# GENERALIZATION OF THE VISCOUS STRESS TENSOR TO THE CASE OF NON-SMALL GRADIENTS OF HYDRODYNAMIC VELOCITY: A PATH TO NUMERICAL MODELING OF TURBULENCE NON-LOCALITY

A.B. Kukushkin[1,2]

[1]*National Research Center "Kurchatov Institute", Moscow, 123182, Russian Federation*
[2]*National Research Nuclear University MEPhI, Moscow, 115409, Russian Federation*

E-mail address: Kukushkin_AB@nrcki.ru, kukushkin.alexander@gmail.com

**Abstract.** Generalization of the Chapman-Enskog method to the case of large gradients of hydrodynamic velocity allowed us to obtain an integral (over spatial coordinates) representation of the viscous stress tensor in the Navier-Stokes equation. In the case of small path lengths of the medium disturbance, the tensor goes over to the standard form, which, as is known, is difficult to apply to the description of tangential discontinuities and separated flows. The obtained expression can allow numerical modeling of the nonlocality of turbulence, expressed by the empirical Richardson $t^3$ law for pair correlations in a turbulent medium.

## 1. Introduction

In modeling hydrodynamic turbulence, the Navier-Stokes equation is used for the dynamics of the mean mass velocity (hydrodynamic velocity)

$$\rho\left(\frac{\partial v_i}{\partial t} + v_k \frac{\partial v_i}{\partial r_k}\right) = -\frac{\partial P}{\partial r_i} + \frac{\partial \sigma_{ik}}{\partial r_k} + F_i, \qquad (1.1)$$

with viscous stress tensor

$$\sigma_{ik} = \eta\left(\frac{\partial v_i}{\partial r_k} + \frac{\partial v_k}{\partial r_i} - \frac{2}{3}\delta_{ik}\frac{\partial v_j}{\partial r_j}\right). \qquad (1.2)$$

It is this description of hydrodynamics, taken in the special case of an incompressible fluid, that was used to construct the world-famous theory of homogeneous isotropic turbulence in the model of Kolmogorov [1]. The given expression for $\sigma_{ik}$ is obtained by the Chapman-Enskog method [2] from the Boltzmann kinetic equation for the single-particle distribution function of particles (atoms or molecules) in phase space, $f(\boldsymbol{p}, \boldsymbol{r}, t)$. In this case, the coefficient of (dynamic) viscosity η turns out to be inversely proportional to the cross-section $\sigma_{col}$ of pair collisions of particles, contrary to the intuitive expectation of an increase in viscosity with an increase in such a cross-section (see, for example, [3]):

$$\eta \sim \frac{\mu v_T}{\sigma_{col}}, \qquad (1.3)$$

where μ is the reduced mass, $v_T$ is the thermal velocity of the particles. The obvious paradox can be explained by the fact that the Chapman-Enskog method actually uses the condition of smallness of the gradients of hydrodynamic quantities (including velocity), which will be more feasible the larger $\sigma_{col}$.

The inapplicability of the standard tensor (1.2) to practical problems of describing the flow around solids and tangential discontinuities in the case of arbitrary flows has long been recognized in the most successful mathematical models. For example, in the widely used Spalart-Allmaras model [4], instead of the viscosity coefficient, a complex function of coordinates and time is used, obtained by solving a differential equation. These models are phenomenological, since their direct derivation from first principles is absent in the available scientific literature, and the structure of the equations indicates the essential role of linking the models to certain specific numerical calculations. This, however, does not prevent the fairly wide application of these models.

Existing mathematical models of turbulence, including various versions of the RANS approach dating back to [5], are local in the sense that they are described by differential equations (more precisely, any integral representation of them reduces to differential equations). An alternative to this is the proposal of Shlesinger et al. [6] to use non-local (i.e., integral over spatial coordinates) models that are not reducible to differential ones and belong to the class of transport models of the "Lévy flights" or "Lévy walks" type (this generalizes Lévy flights to the case where the finite velocity of carriers is taken into account, see review [8]).

An important stimulus for the emergence of the idea [6] was that the nonlocality (superdiffusivity) of turbulence is expressed in the empirical Richardson $t^3$ law [9] for the mean square of the mutual separation of a pair of particles $r_{pair}(t)$ in a liquid or gaseous medium:

$$r_{pair}^2(t) \propto t^3, \qquad (1.4)$$

which differs significantly from such a correlator for ordinary Brownian diffusion, $r^2 \sim t$.

The idea of [6] was used in [10] to construct a phenomenological model of plasma density fluctuation transfer in the plasma turbulence regime. The mathematical apparatus of the model [10] is close to the strictly substantiated Biberman-Holstein model [11, 12], widely used (see [13, 14]) for radiation transfer in spectral lines of atoms and ions in gases and plasmas in the case of complete frequency redistribution in the act of absorption and subsequent emission of a photon by an atom or ion. With the model [10], using the results of the development of the theory of non-local transport in [15-19], it was shown that the plasma density fluctuations observed in the T-10 tokamak using cross-correlation reflectometry of EM waves [20, 21] may have a turbulent origin: the decay rate in the Lévy distribution for the probability of the free path of density fluctuations across a strong magnetic field (i.e., along a small radius of a toroidal plasma column), found by solving the inverse problem from the spectral and radial dependence of the measured cross-correlation function, turned out to be close to its analogue, corresponding to the empirical Richardson $t^3$ law[9].

In [22] an attempt was made to look at the model [10] through the eyes of the general ideology [1] as a way of making non-dimensional the characteristic parameters of the problem and subsequent analysis. This property of the approach [1] allows, as indicated in [23], to interpret it as "dimensional reasoning". Indeed, as far as we know, it has not yet been possible to obtain the main results of [1] from the Navier-Stokes equation itself by directly solving this equation. The use of the results of [10] in [22] made it possible to indicate the connection between the results of numerical modeling [26] of turbulence within the framework of the Navier-Stokes formalism and the analytical results [19] for the Green's function of the transport problem in the Lévy walk regime.

It should be noted that the transition in [22] from the transfer of density disturbances in plasma to non-plasma hydrogasdynamics within the framework of the phenomenological model [10] is an increase in the degree of phenomenology, since in plasma there are well-studied mechanisms of energy transfer by emission and absorption of electromagnetic waves, for example, Cherenkov radiation of longitudinal waves in plasma and the inverse process – Landau damping (see examples of non-local heat transfer by Bernstein waves in [24, 25]). In hydrogasdynamics, the process of non-local transfer as a set of acts of birth ("emission") of a localized vortex, its long flight and stop or decay ("absorption"), in our opinion, has not yet been traced to the same extent as the indicated mechanisms of transfer in plasma.

Finally, we note that examples of dissatisfaction of researchers with the Navier-Stokes model include complete neglect of viscosity in many successful examples of numerical modeling (see, for example, the Introduction in [27]). Here, we should also mention the demonstration in [28] of the possibility of smoothing tangential discontinuities only in the case of asymptotically small viscosity.

The movement along the above-described thematic trajectory [15-19, 10, 22] to a non-local theory of hydrodynamic turbulence suggests the need to construct a mathematical model from first principles, including (a) an analysis of the possibility of obtaining a non-local, i.e. integral over the spatial coordinate, description of the viscous stress tensor, and (b) overcoming the above-mentioned paradox of the inverse dependence of the viscosity coefficient on the scattering cross-section (see (1.3)). This is exactly what is done in this article.

## 2. Generalization of the viscous stress tensor

Let us try to generalize the viscous stress tensor by abandoning the condition of smallness of the velocity gradient in the standard derivation of the tensor in the Chapman-Enskog method. We will leave unchanged the first step in this approach, which consists in the assumption of such dominance of pair collisions that the distribution function in the phase space $f(\boldsymbol{p}, \boldsymbol{r})$ will be considered thermodynamically equilibrium, namely, the Maxwellian distribution over velocities in a coordinate system moving with the average mass velocity of the medium (for simplicity, we will restrict ourselves here and below to the case of a single-sort gas):

$$f^{(0)}(\boldsymbol{p}, \boldsymbol{r}, t) = \frac{n}{(2\pi mT)^{3/2}} exp\left(-\frac{(\boldsymbol{p} - m\boldsymbol{v_0}(\boldsymbol{r}, t))^2}{2mT}\right), \qquad (2.1)$$

where $\boldsymbol{p} = m\boldsymbol{v}$ is the momentum of the particles, $m$ is the mass, $\rho = mn$ is the mass density, $n$ is the particle density,

$$n(\boldsymbol{r}, t) = \int f(\boldsymbol{p}, \boldsymbol{r}, t) d\boldsymbol{p}, \qquad (2.2)$$

the average mass velocity has the standard form:

$$v_0(r, t) = \frac{1}{n} \int v f(p, r, t) dp. \tag{2.3}$$

Equation for $\varphi(p, r)$, the correction to $f^{(0)}$ in the full distribution function

$$f(p, r, t) = f^{(0)}(p, r, t)(1 + \varphi(p, r, t)), \tag{2.4}$$

will be presented, on the one hand, in a simplified form, replacing the Boltzmann collision integral with the simplest form (the deviation of the distribution function from the thermodynamic equilibrium state), but, on the other hand, we will leave the derivatives of the function φ on the left-hand side, since against the background of the derivatives of the relatively smooth function $f^{(0)}$ the derivatives of φ can be much larger in the case of the occurrence of shear flows, which can be localized vortices. Thus we obtain the equation:

$$f^{(0)} \frac{\partial v_{0i}}{\partial r_k} \left( V_i V_k - \frac{1}{3} \delta_{ik} V^2 \right) \frac{m}{T} + f^{(0)} \left( \frac{\partial \varphi}{\partial t} + v_k \frac{\partial \varphi}{\partial r_k} + F_k \frac{\partial \varphi}{\partial p_k} \right) = -f^{(0)} \frac{\varphi}{\tau}, \tag{2.5}$$

where the velocity relative to the average mass velocity of the medium is introduced,

$$V = v - v_0(r, t), \tag{2.6}$$

and the choice of relaxation time τ will be discussed below. In (2.5) it is taken into account that the density and average velocity satisfy the continuity equation and the Euler equation:

$$\frac{\partial \rho}{\partial t} + \frac{\partial}{\partial r_k}(\rho v_k) = 0, \tag{2.7}$$

$$\rho \left( \frac{\partial v_{0i}}{\partial t} + v_{0k} \frac{\partial v_{0i}}{\partial r_k} \right) = -\frac{\partial P}{\partial r_i} + F_i, \tag{2.8}$$

where $P$ is the pressure, and $F$ is the vector of the external force density, and the summation is performed over the repeating indices. It is easy to verify that the use of the simplified collision integral in (2.5) in the Chapman-Enskog method leads to a qualitatively close result for the viscosity coefficient η (only the dependence on the temperature T differs due to the difference in the functionals in averaging $\sigma_{col}$ over the particle velocities).

Since we consider φ as the contribution of localized carriers, we will assume that the external forces acting on them can be taken into account approximately by considering the dynamics of carriers against the background of the main medium described by equations (2.7) and (2.8). Then (2.5) can be represented as:

$$\left( \frac{\partial \varphi}{\partial t} + V_k \frac{\partial \varphi}{\partial r_k} \right) = -\frac{\partial v_{0i}}{\partial r_k} \left( V_i V_k - \frac{1}{3} \delta_{ik} V^2 \right) \frac{m}{T} - \frac{\varphi}{\tau}. \tag{2.9}$$

This equation is a typical kinetic equation of transfer, in which there is a source dependent on the gradient of the average mass velocity, and a sink described by the absorption coefficient $1/l$, where $l$ is the characteristic free path:

$$l(\boldsymbol{v}, \boldsymbol{r}, t) = V\tau(V, \boldsymbol{r}, t). \tag{2.10}$$

The solution of this transport equation has a known form, corresponding to the account of the delay of the carriers (see, for example, Chapter 2 in [29]). We will, however, assume that the movement of the carriers relative to the main medium is significantly faster than the movement of the main medium itself. Then we obtain the following expression for $\varphi$:

$$\varphi(\boldsymbol{p}, \boldsymbol{r}, t) = -\int_{r_b}^{r} \frac{\partial v_{0i}}{\partial r_k} \left(V_i V_k - \frac{1}{3}\delta_{ik}V^2\right) \frac{m}{T} \exp\left(-\int_{r'}^{r} \frac{(\boldsymbol{n}_V, d\boldsymbol{r}'')}{V\tau(V, \boldsymbol{r}'')}\right) \frac{(\boldsymbol{n}_V, d\boldsymbol{r}')}{V}, \tag{2.11}$$

where integration over the carrier trajectories is carried out from the boundary of the medium, designated $\boldsymbol{r}_b$.

Next, one can obtain the viscous stress tensor in a standard way:

$$-\sigma_{ik}(\boldsymbol{r}, t) = \int \varphi(\boldsymbol{p}, \boldsymbol{r}, t) f^{(0)}(\boldsymbol{p}, \boldsymbol{r}, t) m V_i V_k d\boldsymbol{p}. \tag{2.12}$$

The integration over the trajectories in (2.11) can be transformed into integration over the volume of the medium if we consider the trajectories of the carriers to be straight from the point of birth to the point of absorption (stopping and possible dissipation). This gives:

$$\sigma_{ik} = \frac{4P}{v_T} \int_0^\infty \frac{x^5 dx}{\pi^{3/2}} e^{-x^2} \int \frac{d\boldsymbol{r}'}{|\boldsymbol{r}-\boldsymbol{r}'|^2} n_i n_k \frac{\partial v_{0j}}{\partial r_l}\left(n_j n_l - \frac{1}{3}\delta_{jl}\right)\frac{m}{T} \exp\left(-\int_{r'}^{r} \frac{(\boldsymbol{n}, d\boldsymbol{r}'')}{l(x, \boldsymbol{r}'')}\right), \tag{2.13}$$

where the integration is carried out over the entire volume of the medium, and the unit vector $\boldsymbol{n}$ corresponds to the solid angle of integration over the space coordinate. In the limit of small path lengths, the integral can be calculated and for the viscous stress tensor we arrive at the standard form (1.2), in which the viscosity coefficient is equal to

$$\eta = P\tau. \tag{2.14}$$

It follows that when choosing the characteristic relaxation time $\tau$, it may be appropriate to take it such that in the limit of small gradients the result (2.14) coincides with the corresponding result of the theory with an exact collision integral. However, the question of choosing $\tau$ can be the subject of a subsequent more detailed analysis.

It is important to note that the resulting correction to the main distribution function does not violate the conservation of the total number of particles in the medium, since

$$\int \varphi(\boldsymbol{p}, \boldsymbol{r}, t) f^{(0)}(\boldsymbol{p}, \boldsymbol{r}, t) d\boldsymbol{p} d\boldsymbol{r} = 0. \tag{2.15}$$

This corresponds to the dynamic exchange of particles between the main medium and the ensemble of carriers of disturbances of this medium.

## 3. Conclusions

1. For a realistic description of turbulence in the Navier-Stokes equation, an operator is needed that adequately describes tangential discontinuities and separation flows, in which, by definition, the derivatives of the hydrodynamic velocity are not limited by anything. The standard viscous stress tensor $\sigma_{ik}$ obtained by the Chapman-Enskog method cannot be such an operator, as can be easily verified it was derived under the assumption of small velocity gradients. Thus, the standard form of $\sigma_{ik}$ is inapplicable to the description of turbulence. This has long been known in numerical modeling of tangential discontinuities and separated flows, but has not yet been realized for homogeneous isotropic turbulence.

2. The nonlocal, i.e. integral over spatial coordinates, expression for the viscous stress tensor obtained in this work has a structure containing a source of excitation carriers of the medium (in the form of the gradient of the hydrodynamic velocity of the medium) and a sink of carriers in the form with an absorption coefficient (the inverse length of the free path). This form is characteristic of the theory of nonlocal transfer, an example of which is the Biberman-Holstein model for the transfer of resonant radiation in plasmas and gases.

3. The obtained expression for $\sigma_{ik}$ resolves the paradox of the Chapman-Enskog theory, in which the viscosity coefficient is inversely proportional to the cross section of pair collisions of particles. Such a dependence is justified only under the condition of a small velocity gradient, which is obviously impossible for long-running localized vortices, which are the most probable carriers of disturbances of the hydrogasdynamic medium, which are capable of implementing the empirical Richardson $t^3$ law for pair correlations in liquids and gases.

4. The obtained expression for the viscous stress tensor is a step towards a non-local theory of turbulence in the sense that the demonstration of the indicated Richardson law requires adequate numerical modeling using the obtained expression for $\sigma_{ik}$.

5. A similar generalization of turbulent hydrogasdynamics can be obtained for heat transfer in gases and liquids, as well as for the hydrodynamics of more complex media, including plasma, since the Chapman-Enskog method is also used to obtain the plasma hydrodynamics equation (see Braginskii's two-fluid magnetohydrodynamics [30], in which the viscosity coefficient for particle motion along a magnetic field is inversely proportional to the Coulomb logarithm).

6. The presented method can be considered as a way to explain the reasons why Kolmogorov's brilliant mathematical method [1], which has not yet been derived from the Navier-Stokes equation by directly solving this equation, has been successful in predicting or explaining many experiments.

**Acknowledgments.** The author is grateful to all co-authors of the cited publications.